\documentclass{aa}
\usepackage[breaklinks,colorlinks,linkcolor=blue,citecolor=blue,urlcolor=blue,hyperfootnotes=true]{hyperref}
\usepackage{graphicx}
\usepackage{txfonts}
\usepackage{url}
\usepackage{graphicx}

\usepackage{xcolor}
\usepackage[normalem]{ulem}

\begin{document}

\title{Evolution of the local spiral structure of the Milky Way revealed by open clusters}

\author{C. J. Hao\inst{1,2}, Y. Xu\inst{1,2}, L. G. Hou\inst{3}, S. B. Bian\inst{1,2}, J. J. Li\inst{1}, Z. Y. Wu\inst{3,4},
Z. H. He \inst{1,2}, Y. J. Li\inst{1}, D. J. Liu\inst{1,5}}

\institute{Purple Mountain Observatory, Chinese Academy of Sciences, Nanjing 210023, PR China
\email{xuye@pmo.ac.cn}
\and
School of Astronomy and Space Science, University of Science and Technology of China, Hefei 230026, PR China
\and
National Astronomical Observatories, Chinese Academy of Sciences, 20A Datun Road, Chaoyang District, Beijing 100101, PR China
\and
School of Astronomy and Space Science, University of Chinese Academy of Sciences, Beijing 101408, PR China
\and
College of Science, China Three Gorges University, Yichang 443002, PR China
}

\date{Received 19 February 2021 / Accepted 27 June 2021}
\titlerunning{Evolution of the Galactic spiral structure}
\authorrunning{Hao et al}

\abstract{
The structure and evolution of the spiral arms of our Milky Way are
basic but long-standing questions in astronomy. In particular, the
lifetime of spiral arms is still a puzzle and has not been well
constrained from observations.
In this work, we aim to inspect these issues using a large
catalogue of open clusters. We compiled a catalogue of 3794 open
clusters based on {\it Gaia} EDR3. A majority of these clusters have accurately
determined parallaxes, proper motions, and radial velocities.
The age parameters for these open clusters are collected from
references or calculated in this work.
In order to understand the nearby spiral structure and its
evolution, we analysed the distributions, kinematic properties,
vertical distributions, and regressed properties of subsamples
of open clusters.
We find evidence that the nearby spiral arms are compatible with a
long-lived spiral pattern and might have remained approximately
stable for the past 80 million years.
In particular, the Local Arm, where our Sun is currently located, is
also suggested to be long-lived in nature and  probably a major arm
segment of the Milky Way.
The evolutionary characteristics of nearby spiral arms show that the
dynamic spiral mechanism might be not prevalent for our
Galaxy. Instead,  density wave theory is more consistent with the
observational properties of open clusters.
}

\maketitle
%
\section{Introduction}

Accurately revealing the spiral structure of the Milky\ Way and its
evolution has attracted the attention of many astronomers.
After the M\,33 galaxy was explored by \cite{hubble1926}, it was
speculated that our Milky Way is also possibly a spiral galaxy.
Since then, much effort has been dedicated to uncover the spiral 
structure of the Galaxy with different kinds of tracers and
methods~\citep[e.g.,][]{lindblad1927, oort1958, Becker1963,
  Becker1964, Becker1970, georgelin1976, Moffat1979, russeil2003,
  paladini2004, dias2005, xu2006, Moitinho2006, Vazquez2008, reid2009,
  Moitinho2010, hou2014, hou2015, reid2019, xu2018, xu2021,
  Poggio2021}. However, because the Sun is deeply embedded in the Galactic
disc, multiple structures at different distances along the
line of sight are superimposed; this makes very difficult to accurately
decompose these structures and depict the actual spiral structure
\citep{xu2018b}.
Moreover, another big challenge is reliably tracing the Galaxy's morphology to the past to
ultimately understand its evolutionary history.

Open clusters (OCs) are one of the good spiral tracers and
have some specific advantages for understanding the
properties of spiral structure of the Galaxy.
An open cluster is a group of stars that formed from the same giant
molecular cloud, which are roughly the same age.
Unlike the other good tracers for investigating the spiral
arms of the Galaxy  (e.g., high-mass star formation region (HMSFR) masers, massive
O--B type (OB) stars, and HII regions), the ages of OCs cover a wide
range, from a few million years (Myr) to tens of billions of years, making
them potentially good tracers for investigating the spiral structure
outlined by young to old objects, and in particular the evolution of
spiral arms.
In addition, the large number of cluster members makes it possible to
derive more accurate values of cluster parameters (e.g., distances,
proper motions, and radial velocities) than those of individual
star. The spiral structure of the Galaxy has been
explored by many research works using OCs as tracers.

The relationship between OCs and spiral arms was first discussed by
\cite{Becker1963, Becker1964}, who studied a sample of 156 OCs with
photometric distances at that time.
Becker suggested that the distribution of those OCs with earliest
spectral type of O--B2 followed three spiral arm segments in the solar
neighbourhood.
In comparison, the distribution of older OCs with earliest
spectral type from B3 to F did not present arm-like structures
and seemed to be random.
These results were confirmed by \cite{Becker1970} and
\cite{Fenkart1979} with larger samples of OCs.
While a diferent explanation was proposed by \citet{janes1982} and
\citet{lynga1982}, these authors suggested that the distributions of OCs seemed
more like some clumpy-like concentrations or complexes
rather than a spiral pattern.
\citet{dias2005} used a sample of 212 OCs to showed that the young OCs
(ages $<1.2\times10^7$ yr) still remain in the three major arm
segments in the solar neighbourhood, that is, the Perseus, Local, and
Sagittarius-Carina Arms. The OCs  $\sim$ 20~Myr in age are
leaving the spiral arms and moving to the inter-arm regions, while for
the older OCs, their distributions do not present clear clump-like or
spiral-like structures.
\citet{dias2005} also estimated the spiral pattern rotation speed of
the Milky Way with OCs and confirmed that a dominant fraction of the
OCs are formed in spiral arms. These results were recently
updated by \citet{dias2019} with the second data release of Gaia
\citep{gaia2018}.
\citet{Moitinho2006} analysed the stellar photometry data for many
OCs; a sample of 61 OCs was used to trace the disc structure in the
third Galactic quadrant.
The detailed spiral structure in the third Galactic quadrant was also
explored by \citet{Vazquez2008} by taking advantage of the data of
young OCs, blue plumes, and molecular clouds.
For a review about the previous efforts to use OCs to study the
Galactic structure, we recommend \citet{Moitinho2010}.
Later, \citet{cama13} studied a sample of young OCs, which are related
with the Perseus Arm and even the Outer Arm \citep[also
see][]{bobylev2014}.
By taking advantage of the second data release of {\it Gaia},
\citet[][]{cantat2018,cantat2019,cantat2020a} have derived the members
and mean parameters for more than 1800 Galactic OCs. The projected
distribution of these OCs onto the Galactic disc was
compared with a spiral arm model of our Galaxy \citep[][]{reid2014}.
%

In brief, it is commonly accepted that young OCs can be used to trace
the nearby spiral arms, while older OCs have more scattered
distribution, which may be used to represent the distribution
of older stellar components in the Galactic disc.
However, the evolutionary properties and the lifetime of 
  the spiral arms of the Galaxy are still puzzles and have not been well
constrained from observations.
With its specialty, OC serves as good tracer for
better understanding these issues.
Fortunately, the number of OCs and the accuracy of OC parameters
(parallaxes, proper motions, and radial velocities) have been improved
significantly in the past few years, largely as a result of the efforts of
{\it Gaia} (see Table~\ref{table:table1}).
In this work, we aim to compile a large catalogue of OCs from
references and accurately derive their parallax distances, proper
motions, and radial velocities from the latest \textit{Gaia} Early Data
Release 3 (hereafter {\it Gaia} EDR3) \citep{gaia2016, gaia2020}.
This largest catalogue of OCs available to date provides a good
chance to study the above issues about spiral
structure of the Galaxy, not only to map the nearby spiral arms, but also to
understand their evolution, and further, to place observational
constraints on the different evolutionary mechanisms.

This article is organised as follows. In Sect. \ref{sample}, we
describe the sample of OCs compiled in this work. The results
and discussions are presented in Sect. \ref{result}, including the
spiral structure in the solar neighbourhood traced by OCs, the
kinematic properties of OCs, the evolutionary properties of their
scale heights, and a regression analysis. Conclusions are given in
Sect. \ref{conclusion}.

\setlength{\tabcolsep}{0.5mm}
\begin{table}[ht]
\centering
\caption{A list of references for the previously known OCs.}
 \begin{tabular}{ccc} 
\hline
 & OCs &  \\
Reference & Number & Data used  \\ \hline 
\cite{dias2002} & 2167 &  WEBDA$^{(1)}$ et al. \\  
\cite{dias2014} & 1805 &  UCAC4$^{(2)}$\\  
\cite{kharchenko2013} & 3006 &  2MASS$^{(3)}$ \& PPMXL$^{(4)}$ \\  
\cite{schmeja2014} & 139 &  2MASS  \\  
\cite{scholz2015} & 63 &  PPMXL \& UCAC4  \\  
\cite{cantat2018} & 1229 &  \textit{Gaia} DR2  \\  
\cite{cantat2019} & 41 &  \textit{Gaia} DR2  \\ 
\cite{castro2018} & 23 &  \textit{Gaia} DR2 \& TGAS$^{(5)}$ \\  
\cite{castro2019} & 53 &  \textit{Gaia} DR2 \\  
\cite{castro2020} & 582 &  \textit{Gaia} DR2  \\  
\cite{liu2019} & 76 &  \textit{Gaia} DR2  \\  
\cite{he2021} & 74 &  \textit{Gaia} DR2  \\  
\cite{ferreria2020} & 25 &  \textit{Gaia} DR2  \\ \hline  
 \end{tabular}
 \tablefoot{(1) WEBDA: \url{http://obswww.unige.ch/webda/}; 
 (2) UCAC = United States Naval Observatory CCD Astrograph Catalog;
   (3) 2MASS = Two Micron All Sky Survey;
   (4) PPMXL: \url{https://dc.zah.uni-heidelberg.de/ppmxl/q/cone/info};
 (5) TGAS = \textit{Tycho-Gaia} Astrometric Solution.
 }
\label{table:table1}
\end{table}

\section{Sample}
\label{sample}

To better understand the spiral structure of the Galaxy with OCs, an
updated OC catalogue with accurately measured parameters from {\it
 Gaia} EDR3 is necessary.
The recently published {\it Gaia} EDR3 contains the positions,
parallaxes, and proper motions for more than 1.5 billion stars of
different types, ages, and evolutionary stages in the whole sky
\citep{gaia2020}. For the sources with magnitude $G<$ 15, the parallax
uncertainties could be 0.02--0.03~mas, and the proper motion
uncertainties could be 0.02--0.03~mas~yr$^{-1}$. {\it Gaia} EDR3
provides excellent database to improve the parameter accuracies for a
large number of OCs.

\begin{figure*}
\centering
\includegraphics[scale=0.30]{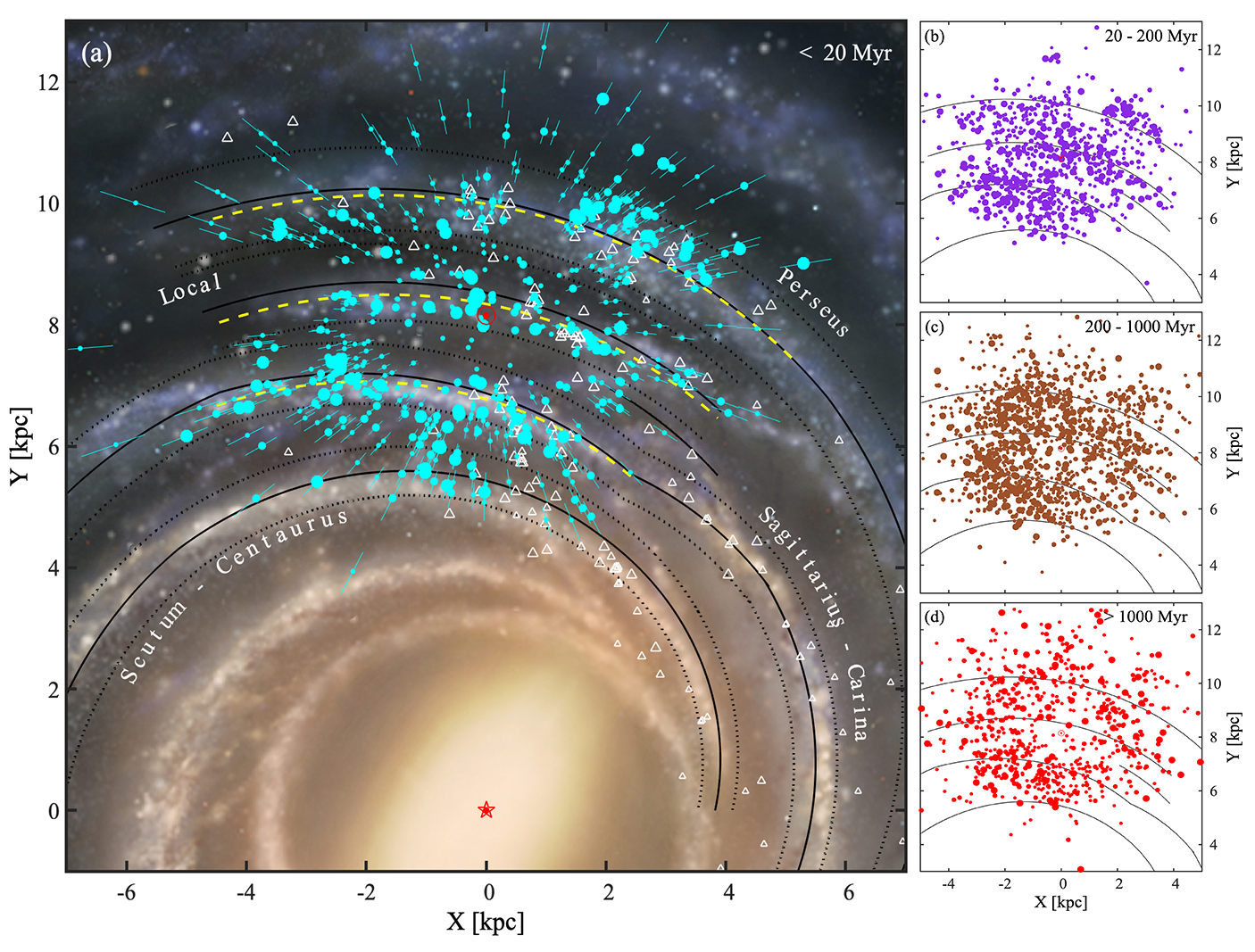}
\caption{Distributions of Galactic OCs. (a) YOCs (cyan dots;
  ages $<$ 20~Myr) and HMSFR masers (white triangles) projected onto
  the Galactic disc. The dot size is proportional to the
  number of cluster members. The solid curved lines trace the centres
  (and dotted lines the widths enclosing 90\% of the masers) of the
  best-fit spiral arms given by \cite{reid2019}. The distance
  uncertainties of the masers are indicated by the inverse size of the
  symbols. The yellow dashed lines denote the best-fit spiral
  arm centres from the distribution of YOCs. The Galactic centre
  (red star) is at (0, 0) kpc and the Sun (red symbol) is at (0, 8.15)
  kpc. The background is a new concept map of the Milky Way credited by
  Xing-Wu Zheng \& Mark Reid BeSSeL/NJU/CFA. The distributions of the
  older clusters are shown in panel (b), (c), and (d) for the OCs
  with ages of 20--200~Myr (purple dots), 200--1000~Myr (brown dots),
  and $>$ 1000~Myr (red dots), respectively. The dot size is
  also proportional to the number of cluster members. The parallax
  uncertainties of the OCs shown are all smaller than 10\%.  }
\label{fig:fig1}
\end{figure*}

\setlength{\tabcolsep}{2.5mm}
\begin{table*}[ht]
\centering
\caption{Best-fit parameters of nearby spiral arms using
  HMSFR masers and OCs of different age groups.}
 \begin{tabular}{ccccccccc} 
\hline
&& Arm & $\varphi$ & $R_{\rm ref}$ & $\beta_{\rm ref}$ & Arm width &&\\ 
&& &  (deg) &   (kpc) &   (deg) & (kpc) &&\\  \hline 
&& Sagittarius-Carina & 17.1 $\pm$ 1.6 & 6.04 $\pm$ 0.09  & 24 & 0.27 $\pm$ 0.04 &&\\ 
&HMSFR masers& Local & 11.4 $\pm$ 1.9 & 8.26 $\pm$ 0.05 & 9 & 0.31 $\pm$ 0.05 &&\\
&& Perseus & 10.3 $\pm$ 1.4 & 8.87 $\pm$ 0.13 & 40 & 0.35 $\pm$ 0.06 &&\\ \hline
&& Sagittarius-Carina & 16.2 $\pm$ 0.4 & 6.72 $\pm$ 0.07 & 2  & 0.27 $\pm$ 0.02 &&\\
&YOCs ($<$ 20~Myr)& Local & 10.8 $\pm$ 0.6 & 8.18 $\pm$ 0.06 & 6  & 0.24 $\pm$ 0.06 && \\ 
&& Perseus & 9.6 $\pm$ 0.8 & 9.42 $\pm$ 0.04 & 20 & 0.33 $\pm$ 0.03 &&\\ \hline
&& Sagittarius-Carina & 15.7 $\pm$ 0.5 & 6.80 $\pm$ 0.07 & 2  & 0.30 $\pm$ 0.03 &&\\
&OCs (20--60~Myr)& Local & 10.7 $\pm$ 0.8 & 8.24 $\pm$ 0.07 & 6 & 0.31 $\pm$ 0.03  &&\\ 
&& Perseus & 9.0 $\pm$ 1.0 & 9.45 $\pm$ 0.06 & 20 & 0.36 $\pm$ 0.04&&\\ \hline
&& Sagittarius-Carina & 15.2 $\pm$ 0.5 & 6.79 $\pm$ 0.07 & 2  & 0.33 $\pm$ 0.04 &&\\
&OCs (60--100~Myr) & Local & 8.9 $\pm$ 0.9 & 8.09 $\pm$ 0.06 & 5 & 0.40 $\pm$ 0.03  &&\\ 
&& Perseus & 9.0 $\pm$ 0.9 & 9.43 $\pm$ 0.07 & 20 & 0.39 $\pm$ 0.05&&\\  \hline 
 \end{tabular}
 \tablefoot{
 For each spiral arm, the best-fit pitch angle ($\varphi$), initial radius
  ($R_{\rm ref}$), reference Galactocentric azimuth angle
  ($\beta_{\rm ref}$), and arm width are listed. The OCs are divided into
  three subsamples: (1) YOCs ($<$ 20~Myr, 633 clusters); (2) OCs with
  ages from 20 to 60~Myr (334 clusters); and (3) OCs with ages from 60 to
  100~Myr (262 clusters).
 }
\label{table:table2}
\end{table*}

Up to now, thousands of OCs have been identified, but scattered in
references.
\cite{dias2002} assembled a large catalogue of OCs based on many
previous studies, which contains 2167 objects.
\cite{dias2014} gave the proper motions and membership probabilities
for 1805 optically visible OCs.
By taking advantage of near-infrared photometric data of
approximately 470 million objects in Two Micron All Sky Survey
~\citep[2MASS,][]{skrutskie2006} and
proper motions from the PPMXL catalogue~\citep{roser2010},
\citet{kharchenko2013} completed a survey of all previously known OCs
called the Milky Way Star Clusters Catalog (MWSC), which lists
3006 objects, including known OCs and candidates, globular
clusters, associations, and asterisms. Subsequently, the MWSC
catalogue was complemented by 139 new OCs at high Galactic latitudes
\citep{schmeja2014} and 63 new OCs identified by \cite{scholz2015}.
After the publication of \textit{Gaia} DR2 \citep{gaia2018}, the
membership probabilities of $\sim$ 1200 OCs presented in previous
catalogues were recalculated by \cite{cantat2018}. These authors also
discovered 60 new OCs. Besides, the radial velocities of 861 OCs in
\cite{cantat2018} were obtained by \cite{soubiran2018}.
\cite{castro2018} developed a machine-learning approach and found 23
new OCs from {\it Gaia} DR2. Soon after, \citet{castro2019} detected
53 new OCs in the direction of the Galactic anti-centre.
\citet{cantat2019} identified 41 new OCs in the direction of the
Perseus Arm.
Meanwhile, 76 new OCs were reported by \cite{liu2019}.
Recently, 582 new OCs were discovered by \cite{castro2020},
25 new OCs were detected by \cite{ferreria2020},
and 74 new OCs were found by \cite{he2021}, respectively.
%


As the first step of this work, we compiled a large
catalogue\footnote{\url{https://cdsarc.u-strasbg.fr/ftp/vizier.submit//hcjgaia2021_v2/}},
which includes 3794 Galactic OCs collected from
references (see Table~\ref{table:table1}). The procedure is
described as follows.
Firstly, 2469 OCs were selected from \cite{kharchenko2013},
which provides details of member stars for each of the
  OCs. Then, the member stars with probabilities greater than
  60\% were picked out to conduct positional cross-matching with
\textit{Gaia} EDR3 using a matching radius of 1~arcsec.
 Making use of the sample-based clustering
search method \citep{hao2020}, which was an adaptation of the OC
search method presented in \cite{castro2018}, we yielded 2457 OCs.
For the OC catalogue of \cite{dias2002}, we cross-matched it with the
catalogue of \cite{kharchenko2013}, and 315 OCs not listed in
  \citet{kharchenko2013} were obtained. The sample-based clustering
search method was applied to these targets, and we got 309
OCs.
Subsequently, we cross-matched the catalogue of \cite{cantat2018} with
these 2766 OCs and then obtained 153 OCs. For these 153 OCs, only
  the member stars with probabilities greater than 60\% were
  picked out. The same procedure was conducted on the catalogue
of \cite{cantat2019}. Then, we obtained detailed parameters of
658 OCs from the catalogues of
\cite{castro2018,castro2019,castro2020}, including the member
stars.
In addition, we collected 76 new OCs from the catalogue of
\cite{liu2019}, 25 new OCs from \cite{ferreria2020}, and 74 new
OCs from \cite{he2021}, respectively. Finally, we obtained a
total of 3794 OCs. The parameters of these OCs were determined
using {\it Gaia} EDR3.

\textit{Gaia} EDR3 provides parallax uncertainties for almost all of
its stars. To obtain precise parallax for each of these OCs, we only
extracted the member stars whose parallax uncertainties are smaller
than 10\%. Not all of the 3794 OCs have member stars with
parallax uncertainties better than 10\%. Hence, a high-accuracy
subsample of 3611 OCs was extracted, in which, 1742 OCs have radial
velocities.
For these 1742 OCs, the means, standard errors, and
uncertainties of their radial velocities were determined using the
method of \cite{soubiran2018}.
After that, the member stars brighter than \textit{G} = 17~mag were
used to determine the position and proper motions for each
of these OCs. For the bright member stars, the uncertainties are
better than 0.05~mas for the position and 0.07~mas~yr$^{-1}$ for the
proper motions, respectively \citep{gaia2020}.
In terms of photometry, for the stars of \textit{G} = 17~mag,
the associated photometric error is 0.001~mag and 0.012~mag for
  \textit{G$_{BP}$}, and 0.006~mag for
  \textit{G$_{RP}$}~\citep{gaia2020}.
The determined parameters of the OCs, together with the reference(s)
where the OC was selected, all are given in the on-line catalogue of
this work.

Some recent works \citep[e.g.,][]{cantat2018,
cantat2020b,Monteiro2020} suggest that there are some false
positive or non-existing clusters in the previous catalogues.
Most of these controversial or erroneous objects are the alleged old
and high-altitude clusters \citep[][]{cantat2020b}. We checked
the 146 false positive or nonexistent clusters given by
\citet{dias2002}, and we found that 81 of these are not included in
our catalogue.
In this work, our aim is to study the spiral arms in the Galactic
disc, and we primarily adopt relatively young OCs (e.g., ages $<$ 100~Myr). 
For the 65 possibly false positive clusters
suggested by \citet{dias2002} and kept in our compiled catalogue,
only 16 of these are $<$ 100~Myr in age. In comparison, the total number
of OCs $<$ 100~Myr in age in our catalogue is 1229.
Hence, we believe that this problem does not influence the following
analysis results about the spiral structure of the Galaxy.
Besides, the ages for the 2837 OCs in our compiled catalogue
were already given in previous works; these are adopted directly. For
the remaining 957 ($\sim$ 25\%) OCs, most of these ($\sim$ 70\%) are
recently identified OCs based on the \textit{Gaia} data. We
determined their ages and the errors following the methods described
in \cite{liu2019}, \cite{hao2020}, and \cite{he2021}.
The age parameter for each of the 3794 OCs and the corresponding
reference(s) are also given in the on-line catalogue.
%


\section{Results and discussions}
\label{result}

\subsection{Spiral structure in the solar neighbourhood}
\label{spiral structure}

Figure~\ref{fig:fig1} presents the projected distributions of OCs
onto the Galactic disc. Figure~\ref{fig:fig1}(a) shows that
the 633 young OCs (YOCs; i.e., ages $<$ 20~Myr) in our
catalogue are concentrated in distinct structures,
which are almost concordant with the known spiral arms outlined by
HMSFR masers \citep{reid2019} and/or bright OB stars~\citep{xu2018,xu2021}.
A large number of YOCs are located in the Perseus, Local,
Sagittarius--Carina, and Scutum--Centaurus Arms, which seem to
be the nurseries for many OCs.
These arm features traced by YOCs extend as far as $\sim$3 to 8~kpc
along different spiral arms.
These properties are in general consistent with previous results
\citep[e.g., see][]{Becker1970,dias2005,Moitinho2006,Vazquez2008,
  dias2019,cantat2018,cantat2020a}.
In Figure~\ref{fig:fig1}(a), it is also shown that the more populous
OCs (i.e., those with more member stars) seem to be more concentrated
on the major spiral arms than the less populous ones.
In addition, spur-like features indicated by the concentration
of some YOCs are visible in the inter-arm regions, which is
consistent with the results shown using HMSFR masers
\citep[][]{reid2019} and bright OB stars
\citep[][]{xu2018,xu2021}. This all suggests that our Milky Way is not a
pure grand-design spiral, but perhaps a multi-armed spiral
galaxy showing complex substructures \citep{xu2016}.

The distributions of older OCs are shown in
Figure~\ref{fig:fig1}(b), (c) ,and (d). Consistent with previous
results~\citep[e.g.,][]{Moitinho2006, Vazquez2008, bobylev2014,
  Reddy2016, dias2005,cantat2018, cantat2020a}, the OCs are gradually
dispersed in the Galactic disc as they age. Structural features
are still discernible for the populous OCs with ages of tens of
million years, but the older OCs (e.g., ages $>$ 200~Myr) present more
diffuse distributions. Indeed, when using the OCs older than 1000~Myr
as tracers, spiral arm features are almost unrecognisable.
These properties indicate that as the YOCs age they gradually
migrate away from their birthplaces, probably accompanied by the
evolution of spiral arms.
%

\begin{figure*}
\centering
\includegraphics[scale=0.25]{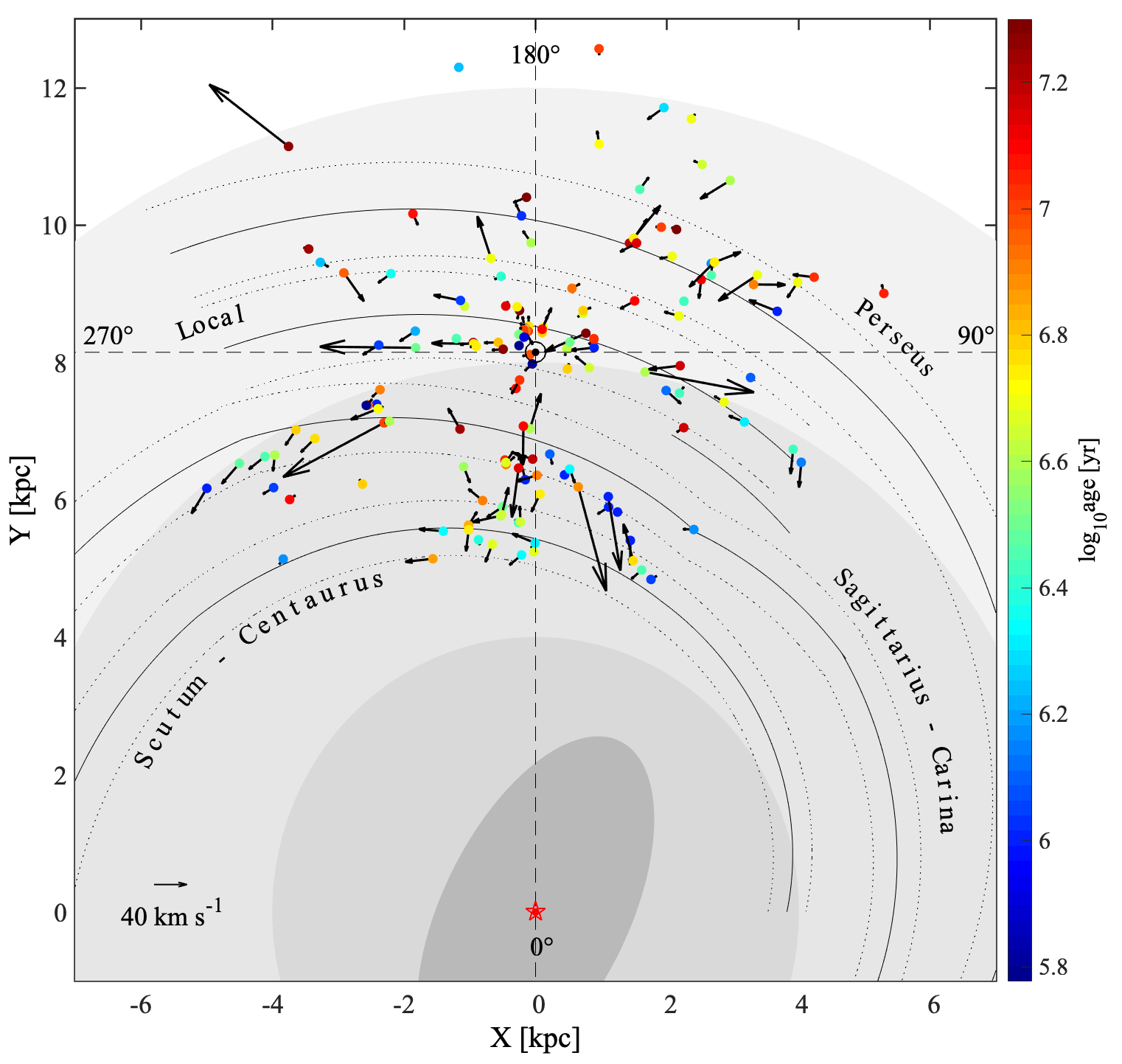}
\caption{Proper motion vectors of the YOCs. A motion scale of
  40~km~s$^{-1}$ is indicated in the bottom left corner of the
  plot. The ages of OCs are colour coded. The background is the same as
  that of \citet[][]{reid2019}.}
\label{fig:fig2}
\end{figure*}

\begin{figure}
\centering
\includegraphics[width=0.48\textwidth]{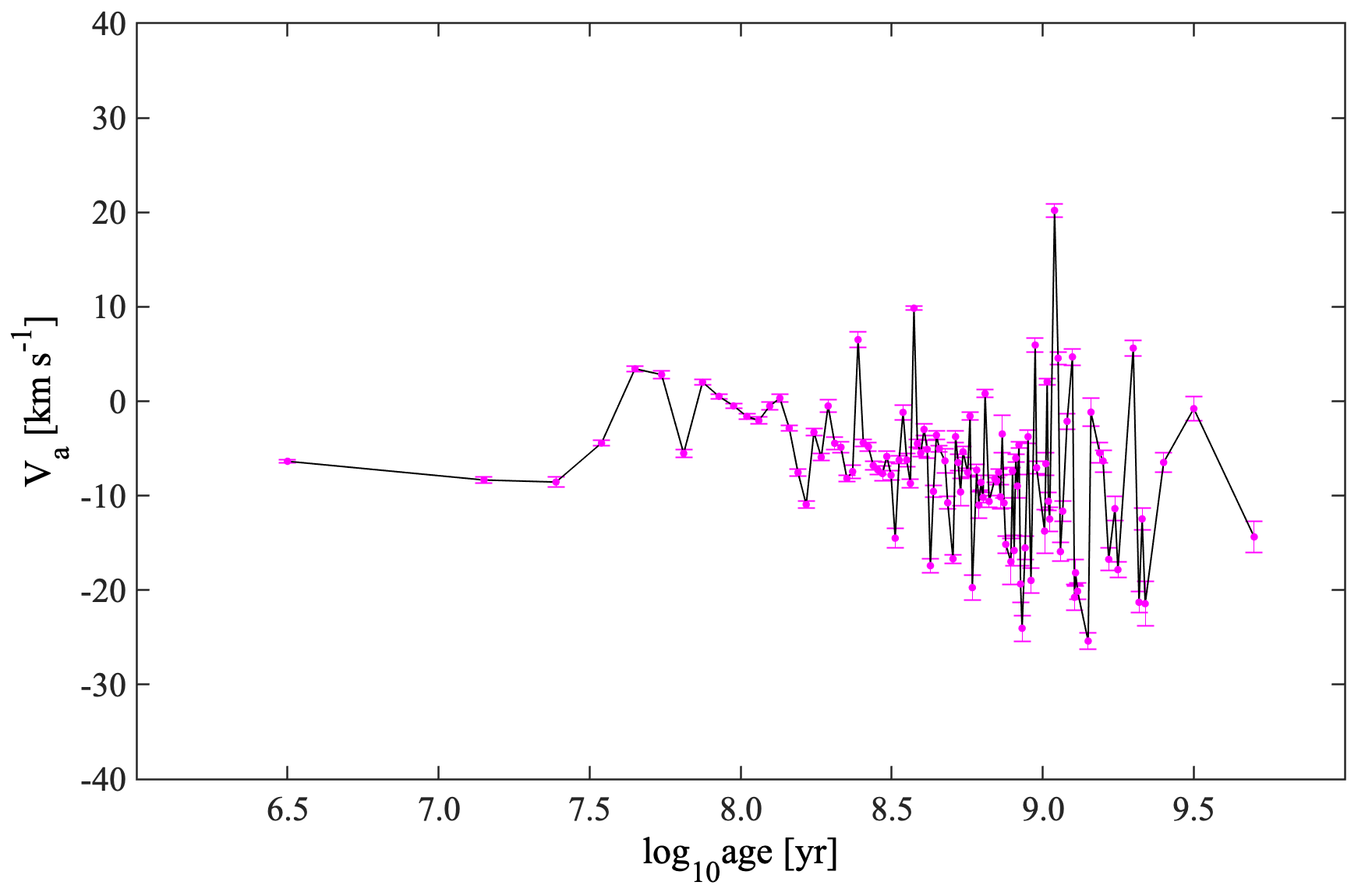}
\caption{Average peculiar motions of the OCs in the direction of
  Galactic rotation as a function of cluster age. The age bin is
  5~Myr. The magenta symbols and error bars indicate the average PMs
  and their associated errors.}
  \label{fig:fig3}
\end{figure}

Inspired by the above phenomena, we considered that as OCs gradually
age the spiral arms traced by OCs of different ages are
probably not the same; instead, they can be used to reflect the
evolution of the spiral arms of the Galaxy.
A subsample of OCs below 100~Myr in age is extracted from our
catalogue and used to explore this issue. To ensure that there are plenty of OCs in each age
interval, we divide those OCs that are
below 100~Myr into three different age groups of $<$
20~Myr (633 clusters), 20--60~Myr (334 clusters), and 60--100~Myr (262
clusters).
From their distributions, the three nearby spiral arms (i.e., the
Sagittarius-Carina, Local, and Perseus Arms) are discernible. For
the more distant Scutum--Centaurus Arm, as fewer OCs are located in it,
this arm cannot be well traced and hence is omitted in the
following analysis.
Then, in order to make a comparison with the maser results, we
determine the parameters of spiral arms with the same method of
\citet[][]{reid2019}, which is used in analysing their
maser parallax data.
To fit the pitch angle and arm width for a spiral arm traced
by YOCs, a logarithmic spiral-arm model is adopted \citep{reid2014},
\begin{equation} 
\\\\
 \ln (R  / R_{\mathrm{ref}}) = -  (\beta - \beta_{\mathrm{ref}}) \tan \varphi,
\end{equation}
where \textit{R}$_{\rm ref}$ is the reference Galactocentric radius,
$\beta_{\rm ref}$ is the reference azimuthal angle, and $\varphi$ is the
pitch angle. The zero point of the Galactocentric azimuthal angle,
$\beta$, is defined as a line towards the Sun from the Galactic centre,
and the azimuthal angle increases towards the direction of Galactic
rotation. The quantity \textit{R} is the Galactocentric radius at a Galactocentric
azimuth $\beta$. The fitting results are listed in
Table~\ref{table:table2}.


In comparison to the values derived by using HMSFR masers, the
pitch angles of the three nearby spiral arms traced by OCs tend to be
smaller; the differences become larger when older OCs were
used as tracers, thus implying a tighter winding of the spiral arms
traced by older OCs.
This is possibly a true feature for the Milky Way; however, the
differences of pitch angles are not significant when considering
the fitting uncertainties and higher quality data of OCs would be
necessary to confirm this.
Besides, arm width is also an important parameter for a spiral arm. 
We note that the fitted arm widths for the OCs with ages of
60--100~Myr are inclined to be larger than those of the HMSFR
masers. In comparison, the arm widths derived from YOCs are in general
consistent with HMSFR masers.

\begin{figure*}[t]
    \centering
   \includegraphics[width=0.70\textwidth]{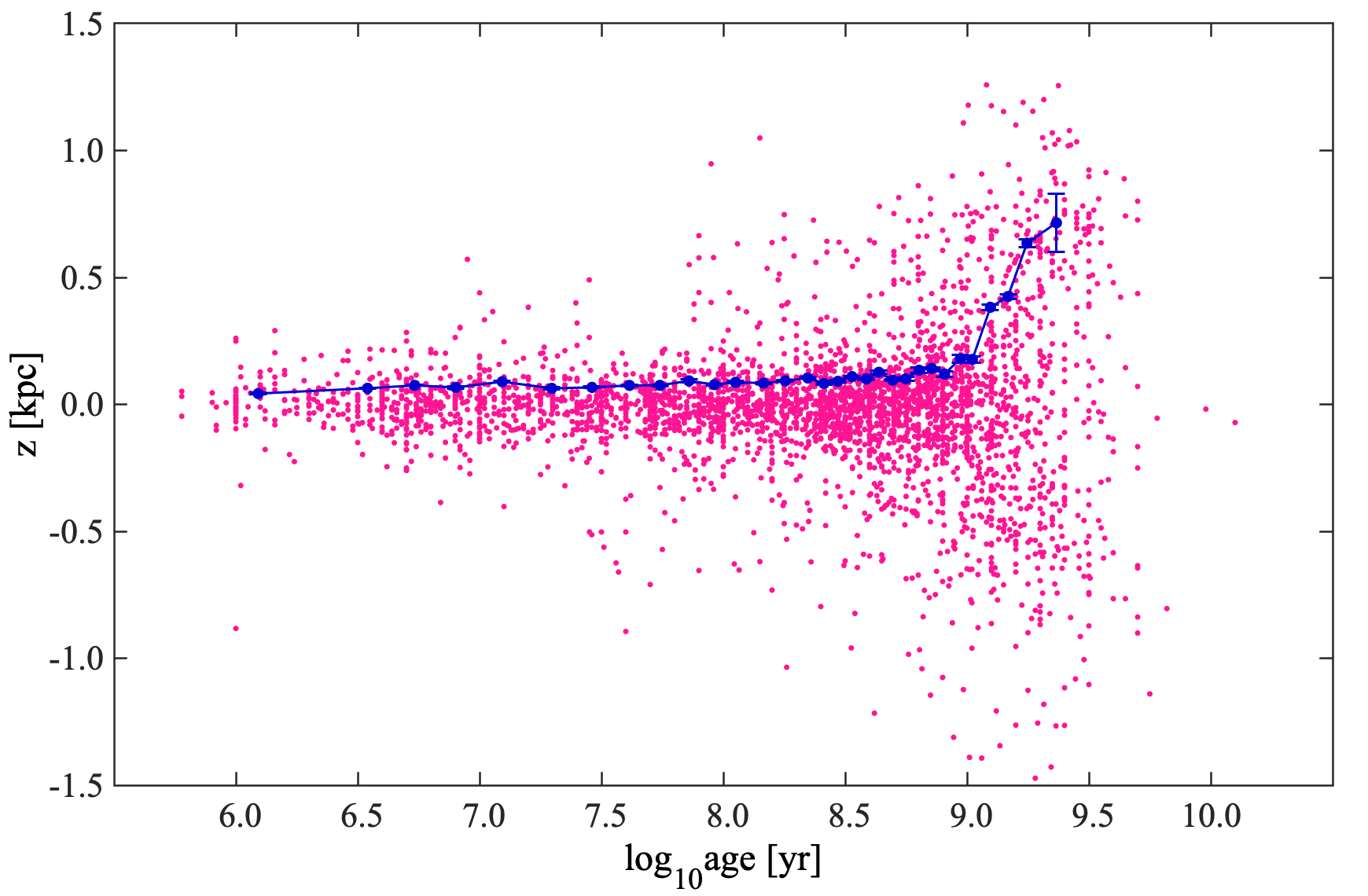}
    \caption{Evolution of the scale height of OCs with OC
      age. The 3376 OCs within 4.0~kpc of the Sun are divided into 32
      bins according to their ages. The blue line indicates the
      evolution of the scale height with cluster age. The blue dots
      correspond to the typical values of log$_{10}$(age) in each
      bin. The solid magenta dots show the distances of the OCs to the
      Galactic plane as a function of cluster age.}
    \label{fig:fig4}
\end{figure*}


\subsection{Kinematic properties of OCs}

With the distances, proper motions, and radial velocities for the
1742 OCs in the high-accuracy sample, we can derive their
three-dimensional (3-D) velocity information following the methods of
\cite{xu2013} and \cite{reid2019}.
The 3-D velocities were straightforwardly calculated using the linear
speeds projected onto the celestial sphere and radial velocities,
which were obtained from the proper motions and distances.
Then, the peculiar motions (PMs; i.e., non-circular motions) of the
OCs were estimated by subtracting the effect of Galactic rotation and
the solar PMs using the updated Galactic parameters.
In this approach, we adopted a Galactic rotation speed of 236 $\pm$
7~km~s$^{-1}$, where the Sun is at a distance of 8.15 $\pm$ 0.15~kpc
to the Galactic centre \citep{reid2019}. Solar PMs of
\textit{U}$_{\odot}$ = 10.6 $\pm$ 1.2~km~s$^{-1}$,
\textit{V}$_{\odot}$ = 10.7 $\pm$ 6.0~km~s$^{-1}$, and
\textit{W}$_{\odot}$ = 7.6 $\pm$ 0.7~km~s$^{-1}$ were assumed
\citep{reid2019}, which are the velocity components towards the
Galactic centre in the direction of Galactic rotation and
towards the North Galactic Pole, respectively.
The sources with an uncertainties larger than 10~km~s$^{-1}$ were
eliminated in the following analysis.

In Figure~\ref{fig:fig2}, we present the PMs of YOCs in the solar
  neighbourhood.
The PMs of YOCs seem to be random. As the gravitational potential of
the Galactic bar is weak in these regions, the PMs of YOCs should be
intrinsic. We find that the average peculiar motion is 1.3
$\pm$ 0.2~km~s$^{-1}$ towards the Galactic centre and $-$6.8 $\pm$
0.2~km~s$^{-1}$ in the direction of the Galactic rotation. Hence,
there is an average lagging motion for YOCs behind the Galactic
rotation. Similar properties were recently noticed for HMSFR masers
\citep[][]{reid2019} and OB stars~\citep{xu2018b}.
We also find that the average lagging motion tends to increase
as the OCs age (Figure~\ref{fig:fig3}).

As discussed in Sect.~\ref{spiral structure}, the YOCs gradually migrate 
away from their birth arms. With the kinematic
data, the mean time for the YOCs to traverse their natal spiral arms
could be estimated.
The average arm width \textit{w} is taken as 0.31 kpc, which is the
mean width of the segments of the Perseus, Local, and
Sagittarius-Carina Arms \citep{reid2019}.
We select the YOCs that are located in spiral arms according to the
best-fit model of \citet{reid2019}, which is consistent with our
fitting results with YOCs as shown in Figure~\ref{fig:fig1} and
Table~\ref{table:table2}.
As shown in Figure~\ref{fig:fig2}, the YOCs approximately migrate
either towards the Galactic centre direction or Galactic anti-centre
direction. Hence, the travel speed of YOCs is defined as the average
of the absolute values of the derived mean velocities.
The travel velocity for the YOCs located in spiral arms, $v$, is
estimated to be 14.6 $\pm$ 0.2~km~s$^{-1}$.
The mean time taken by a YOC to leave its natal spiral arms is simply
estimated as $t = w \div v =$ 20.8 $\pm$ 2.2~Myr.
%


\subsection{Scale height of OCs}
\label{sec:scale height}

In this section, we focus on the vertical distribution of
OCs. The high-accuracy subsample with 3611 OCs (Sect.~\ref{sample}) is
adopted. Statistically, about 43.4\% of the 3611 OCs in our sample
are located above the International Astronomical Union (IAU) defined
Galactic plane ($b=0^\circ$). In previous works, asymmetric
vertical distributions of gas or stars above and below the
IAU-defined Galactic plane were commonly found
\citep[e.g.,][]{ander19,reid2019}, which are generally interpreted as
the Sun slightly deviating from the Galactic midplane towards the
north Galactic pole, that is, the height of the Sun $z_{\odot}$ > 0~pc. Many
efforts have been made to determine the value of $z_{\odot}$
\citep[e.g.,][and references therein]{cantat2020a}.
By using the OCs with accurate {\it Gaia} parallaxes, we can refine
the Galactic plane.
We divide a subsample of 3376 OCs within 4.0~kpc of the Sun into 32
groups according to the OC ages, which are used to determine
\textit{z}$_{\odot}$. Then, the \textit{z}$_{\odot}$ values of the 32
groups are averaged to obtain the mean vertical displacement of the
Sun with respect to the Galactic plane.
Here, the values of \textit{z}$_{\odot}$ and the scale height are
determined using the following function \citep{maiz2001}:
\begin{equation} 
\\\\
N (z) = N \times \exp{\left[-\frac {1} {2}\left(\frac {z - z_{\odot}} { h}\right)^2\right]},
\end{equation}
where $N$ is the central space density of OCs at \textit{z} =
\textit{z}$_{\odot}$ with respect to the Galactic plane ($b=0^\circ$)
and $h$ is the scale height. By analysing the vertical distribution of
OCs with the accurate {\it Gaia} parallaxes, the value of
\textit{z}$_{\odot}$ is refined to be 17.5 $\pm$ 3.8 pc above the true
Galactic plane.
This value is consistent with the results given by \citet{buckner2014}
and \citet{karim2017}, but smaller than that of \cite{cantat2020a}.
The value of \textit{z}$_{\odot}$~=~17.5~$\pm$~3.8~pc is adopted in
this work to adjust the $z$-offsets of the OCs and calculate the
scale heights.

Figure~\ref{fig:fig4} shows the evolution of the scale height with
cluster age for the 3376 OCs within 4.0~kpc of the Sun, which is
consistent with the dispersion of the projected OCs as time goes by.
The scale height is about 42~pc for the OCs with a mean age of $t_{\rm
OC}\sim$ 1.5~Myr and increases to about 130~pc for the OCs with
$t_{\rm OC}\sim$ 850~Myr.
For the OCs with ages greater than $t_{\rm OC}\sim$ 1~Gyr, the scale
heights increase obviously.
For the YOCs (ages $<$ 20~Myr), the scale height is estimated to be
70.5 $\pm$ 2.3~pc. While for the OCs with ages of 20--100~Myr, the
value is 87.4 $\pm$ 3.6 pc.
The deviations of the scale heights of the very young OCs from those
of HMSFR masers \citep[19 $\pm$ 2 pc;][]{reid2019} and O--B$_{5}$
stars \citep[34 $\pm$ 3 pc;][]{maiz2001} are small, but increases
gradually as OC age.
These results are consistent with the properties discussed in
Sect.~\ref{spiral structure}. The OCs within an age of a few million
years are still located in or very close to their birthplaces, hence
have a small scale height. The spiral structure traced by these OCs
resembles that of HMSFR masers or OB stars. But as the OCs
age, the OCs gradually migrated further from their birthplaces and
have larger scale height.
%

\begin{figure*}
\centering
\includegraphics[scale=1.48]{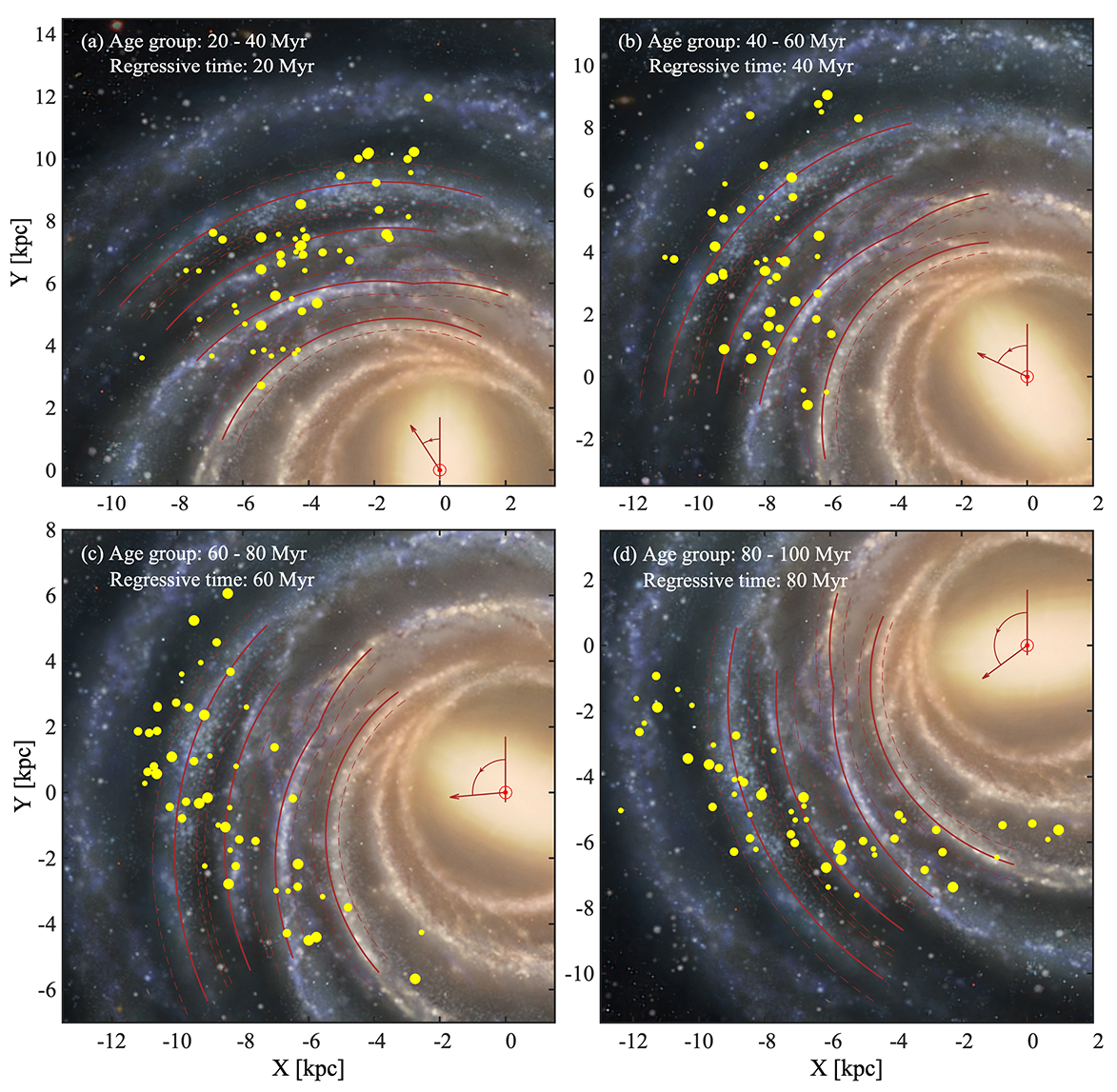}
\caption{Distributions of the regressed OCs. The OC ages have been
  grouped into: (a) 20--40~Myr, (b) 40--60~Myr, (c) 60--80~Myr,
  (d) 80--100~Myr, and traced back to 20, 40, 60, and 80 million years
  ago, respectively. The solid curved lines trace the centres (and
  dotted lines the widths enclosing 90\% of the masers) of the
  best-fit spiral arms given by \cite{reid2019}, and have been
  rigidly rotated back to the past for comparison. The background is a new
  concept map of the Milky Way (credited by: Xing-Wu Zheng \& Mark
  Reid BeSSeL/NJU/CFA), which is also rigidly rotated back to the
  past. The adopted pattern speed is 27~km~s$^{-1}$ kpc$^{-1}$.}
\label{fig:fig5}
\end{figure*}

\begin{table*}[ht!]
\centering
\caption{Best-fit results of the spiral arm parameters for the
  present-day OCs and the regressed OCs.}
\begin{tabular}{cccccccc}
    \hline
&Age groups & 0--20~Myr & 20--40~Myr & 40--60~Myr & 60--80~Myr & 80--100~Myr &\\ 
&Regress to: & & 20 MYA & 40 MYA & 60 MYA & 80 MYA& \\ 
\hline
&& & & Pitch angle: $\varphi$  (degree) & & \\
\hline
&Sgr--Car Arm & 16.2 $\pm$ 0.4 & 16.3 $\pm$ 1.0 & 16.2 $\pm$ 1.1 & 16.2 $\pm$ 1.0 & 16.0 $\pm$ 1.0 &\\
&Local Arm & 10.8 $\pm$ 0.6 & 10.8 $\pm$ 0.9 & 10.5 $\pm$ 1.0 & 10.5 $\pm$ 0.9 & 10.6 $\pm$ 1.0 &\\
&Perseus Arm & 9.6 $\pm$ 0.8 & 9.3 $\pm$ 0.9 & 9.5 $\pm$ 0.9 & 9.6 $\pm$ 0.9 & 9.5 $\pm$ 1.0 &\\ \hline
&& & & Arm width: $w$  (kpc) & & &\\ 
\hline
&Sgr--Car Arm & 0.27 $\pm$ 0.02 & 0.24 $\pm$ 0.09 & 0.24 $\pm$ 0.09 & 0.25 $\pm$ 0.09 & 0.25 $\pm$ 0.10& \\
&Local Arm & 0.24 $\pm$ 0.06 & 0.28 $\pm$ 0.03 & 0.20 $\pm$ 0.05 & 0.23 $\pm$ 0.10 & 0.22 $\pm$ 0.09 &\\
&Perseus Arm & 0.33 $\pm$ 0.03 & 0.29 $\pm$ 0.09 & 0.29 $\pm$ 0.09 & 0.30 $\pm$ 0.10 & 0.31 $\pm$ 0.10 &\\
    \hline
    \end{tabular}
\label{table:table3}
\tablefoot{For each of the spiral arms
  near the Sun, we list its best-fit pitch angle ($\varphi$) and
  arm width (\textit{w}), including their 1$\sigma$ errors.
  Sgr--Car: Sagittarius-Carina Arm; MYA:
    million years ago.}
\end{table*}


\subsection{Regression analysis}

Based on multiwavelength observations, we have some knowledge
about the spiral structure of our Milky Way
\citep[e.g.,][]{xu2018b}. However, the
lifetime of the spiral arms of the Galaxy of the galaxy is still a puzzle; 
the question is whether they are long-lived or
short-lived.
Many of the older OCs (ages $>$ 20~Myr) in our sample have accurate
parallaxes, proper motions, and radial velocities. With an appropriate
model, we may trace their trajectories back tens of millions of years
ago, until they were very young (i.e., ages $<20$~Myr), and resided in
their natal spiral arms in the past.
With this method, the spiral structure in the past could be
explored and compared with its current view.
Such an approach provides a good and special way to inspect the
evolutionary history of nearby spiral pattern and helps to address
questions such as whether the spiral arms remain stable or have they
evolved with time.
Actually, tracing OCs back to their birthplaces has been
proposed by some previous works~\citep[e.g.,][]{dias2005,dias2019}
and used to determine the pattern speed of the Galaxy by assuming a
stable spiral pattern.
In the following, we adopt a regression analysis method to
inspect the stability of nearby spiral structure.

Since the YOCs ($<$ 20~Myr) can well trace the spiral arms, we perform
the regression analysis using 20-Myr intervals. The OCs in the age
ranges of 20--40~Myr, 40--60~Myr, 60--80~Myr, and 80--100~Myr are
regressed to 20, 40, 60, and 80~million years ago, respectively, by
adopting a commonly used model of OC motions in the Milky Way
\citep{wu2009}. This model has been widely used to calculate the
orbits of different objects \citep[e.g.,][]{odenkirchen1997,
  allen2006}, which employs an axisymmetric Galactic
gravitational potential \citep{allen1991}, consists of a spherical
central bulge and a disc, plus a massive, spherical halo extending to
a distance of 100~kpc from the Galactic centre \citep{paczynski1990,
  flynn1996}. The total mass of the model is 9.0 $\times$
10$^{11}$~M$_{\odot}$, and the local total mass density in the solar
neighbourhood is 0.15 M$_{\odot}$ pc$^{-3}$. The results of regression
are shown in Figure~\ref{fig:fig5}. We also fit spiral arms to the
distributions of regressed OCs with the same method described in
Sect.~\ref{spiral structure}. The fitting results are listed in
Table~\ref{table:table3}.

At present, about 72\% of the YOCs are located in the spiral arms
defined by HMSFR masers. In comparison, the percentages are about 39\%
$-$ 59\% for the older clusters in the four different age groups
(20$-$40~Myr, 40$-$60~Myr, 60$-$80~Myr and 80$-$100~Myr).
After regression, many OCs are traced back to spiral arms in the past
as shown in Figure~\ref{fig:fig5}. We found that the percentages of
regressed OCs in spiral arms increase to 71\%$-$85\%, which are
comparable to the present-day value.
The vertical scale height of the present-day OCs in the age groups of
20--100~Myr is about 87 pc. After regression, the scale height
decreases to about 69 pc, close to the value for the present-day YOCs
($\sim$ 71 pc).
In addition, we also notice that the best-fit pitch angles and arm
widths of the Sagittarius-Carina, Local, and Perseus Arms at
different regression times are consistent with each other within the
uncertainties, which are all are very close to the present-day values traced by
YOCs (Table~\ref{table:table3}).
Hence, the spiral-arm features depicted by regressed OCs are roughly
concordant with those traced by present-day YOCs.
These results jointly suggest that the spiral arms near the Sun
probably are compatible with a long-lived pattern, which has
remained stable in the past 80~Myr.
The Local Arm, which was thought to be a spur or branch but now is
suggested as a major spiral arm segment, also has probably existed for
a long time.
However, although the spiral pattern can be roughly recognised over
almost 80 Myr, it is not enough to support a long-lived stable
pattern. It is also possible that the pattern could be in the last
stages of disintegration. In the next works, we expect to
further inspect this issue with more OCs and an improved model of OC
motions in the Milky Way.
The evolutionary characteristics of nearby spiral arms revealed by OCs could 
provide some observational constraints on the formation mechanisms of 
the spiral arms of the Galaxy, which are hotly debated all along. There are two major 
competing scenarios: the dynamic spiral mechanism and the density wave 
theory.
Proponents of the former argue that the spiral arms are short-lived,
transient, and recurrent \citep{toomre1972, sellwood1984}. Besides, the
dynamic spiral mechanism may be able to explain the offsets in some
local regions, but it does not predict the systematic positional
offsets between spiral arms delineated by different kinds of
tracers.
Our above results are compatible with the long-lived and rigid,
rotating nature of nearby spiral arms, and hence seem to conflict with
the predictions of the dynamic spirals, indicating that this
mechanism might be not prevalent for our Milky Way.
In comparison, the (quasi-stationary) density wave theory suggests that the 
spiral arms are long-lived and rigidly rotated; this theory also predicts 
systematic spatial offsets between the spiral arms traced by HMSFRs and 
old stars \citep{lin1964, lin1966, shu2016}.
In this work we show that the evolutionary properties of spiral
arms indicated by OCs are consistent with the expectations of the
density wave theory. Besides, the results from the study addressing
the metallicity gradient step near the Galactic corotation radius
with OCs and cepheids also suggested long-lived spiral arms for the
Milky Way \citep{lepine2003}.
However, we noticed that the density wave theory expects that the
Milky Way has four major spiral arms and the Local Arm should not
exist \citep{yuan1969}.
The presence of the long-lived Local Arm ($\gtrsim$ 80~Myr)
implied by our analysis enhances the challenge previously posed on
this issue \citep{xu2013, xu2016}.
\citet[][]{lepine2017} has interpreted the Local Arm as an
outcome of the spiral corotation resonance, which traps arm tracers
and the Sun inside it \citep[also see][]{baros2021}.

Density wave theory predicts that the Galactic spiral pattern
should remain unchanged over a long time and rotate at a constant
speed \citep{shu2016}. With the regression analysis method, we can
constrain the spiral pattern speed by comparing the OC locations
after regression with the positions of spiral arms rigidly rotated
back to that time.
Based on this, the present-day spiral arms are rotated back to
20, 40, 60, and 80~million years ago, respectively, and then compared with
the distributions of the OCs after regression (Figure~\ref{fig:fig5}).
Under the assumption of a constant pattern speed, which is independent
of the Galactocentric distance, the pattern speed is limited to be
26--28~km~s$^{-1}$~kpc$^{-1}$, which is consistent with the
values given by for example \citet{dias2005, dias2019}, and
\citet{gerhard2011}. The results also suggest that the regression
analysis method used in this work is feasible.


\section{Conclusions}
\label{conclusion}
In summary, we have studied the evolution of the spiral arms of the 
Galaxy using a large catalogue of OCs based on {\it Gaia} EDR3.
A regression analysis method applied to OCs indicates that the spiral
arms near the Sun are compatible with a long-lived spiral
pattern, which might have been essentially stable for at least
the past 80 million years.
In particular, the Local Arm is also probably long-lived in nature, and
well traced by many YOCs, HMSFR masers, and OB stars, all of which
suggest that the Local Arm is probable a major arm segment of the
Milky Way.
These evolutionary characteristics of spiral arms are more consistent
with the expectations of the density wave theory. The dynamic
spiral mechanism might be not prevalent for the Milky Way.
These results are expected to be confirmed by conducting further
studies with more OCs in the Milky Way.


\begin{acknowledgements}
We appreciate the anonymous referee for the instructive
comments which help us a lot to improve the paper. This work was
funded by the NSFC, grant numbers 11933011, 11873019, 11673066,
11988101 and by the Key Laboratory for Radio Astronomy. L.G.Hou thanks
the support from the Youth Innovation Promotion Association CAS. We
acknowledge Dr. Mark J. Reid for allowing us to use his program,
utilised to fit the pitch angles of the Galactic spiral arms. We used
data from the European Space Agency mission \textit{Gaia}
(\url{http://www.cosmos.esa.int/gaia}), processed by the \textit{Gaia}
Data Processing and Analysis Consortium (DPAC; see
\url{http://www.cosmos.esa.int/web/gaia/dpac/consortium}). Funding for
DPAC has been provided by national institutions, in particular the
institutions participating in the \textit{Gaia} Multilateral
Agreement.
\end{acknowledgements}


%
%


\end{document}